\documentclass[a4paper,11pt]{article}

\usepackage{amsthm,amsmath,amssymb,amsfonts,pstricks,pst-node,graphics}

\usepackage{texdraw}

\DeclareMathAccent{\wtilde}{\mathord}{largesymbols}{"65}
\DeclareMathAccent{\what}{\mathord}{largesymbols}{"62}

\newcommand\cS{{\mathcal S}}
\newcommand\cF{{\mathcal F}}

\newcommand\cA{{\mathcal A}}
\newcommand\cB{{\mathcal B}}
\newcommand\cC{{\mathcal C}}
\newcommand\cT{{\mathcal T}}

\newcommand\Z{{\mathbb Z}}
\newcommand\N{{\mathbb N}}
\newcommand\C{{\mathbb C}}

\newcommand\J{{\mathcal I}}
\newcommand\cH{{\mathcal H}}

\newcommand\cM{{\mathcal M}}

\newcommand{\gR}{{\mathfrak{R}}}

\def\im{\operatorname{Im}}

\newcommand\bs{{\bf s}}

\newtheorem{Def}{Definition}
\newtheorem{The}{Theorem}
\newtheorem{Pro}{Proposition}
\newtheorem{Lem}{Lemma}

\makeatletter
\def\wb{\accentset{{\cc@style\underline{\mskip10mu}}}}
\makeatother

\begin{document}
\title{A new recursion operator for the Viallet equation}
\author{Alexander V. Mikhailov$^{\star}$ and Jing Ping Wang$ ^\dagger $
\\
$\star$ Department of Applied Mathematics , University of Leeds, UK\\
$\dagger$ School of Mathematics, Statistics $\&$ Actuarial Science, University of Kent, UK }
\date{}
\maketitle

\begin{abstract}
We present a new recursion  and Hamiltonian
operators for the Viallet equation. This new recursion operator and the
recursion operator found in \cite{mwx1}
satisfy the elliptic curve equation associated with the Viallet equation. 
\end{abstract}

\section{Introduction}
It is well known that integrable equations possess infinitely many generalised
symmetries. These symmetries can be generated by so--called recursion operators
\cite{AKNS74,mr58:25341}, which map a symmetry to a new
symmetry. For example, the
famous Korteweg-de Vries (KdV) equation
$$ u_t=u_{xxx}+6 u u_x$$ possesses a recursion operator
$$\Re=D_x^2+4 u +2 u_x D_x^{-1},
$$
where \(D_x^{-1} \) stands for the right inverse of total derivative \(D_x\).
This is a second order pseudo-differential operator which action is defined on
elements of the space \(\im D_x\). The
infinite hierarchy of symmetries of the KdV equation is generated by
\[
u_{t_j}=\Re^j (u_x),\qquad j=0, 1, 2,\cdots .
\]
Majority of the known recursion operators \cite{wang1} are weakly nonlocal \cite{mr2002g:37093} (Some exceptional ones can be found in \cite{serg7b} and were further studied in \cite{wang09}), that is,
their nonlocal terms are a finite sum of terms of the form
$ K D_x^{-1} \delta_u \rho $, where $K$ is a generalised symmetry and
$\delta_u \rho$ is the variational derivative of a conserved density $\rho$ of the equation.
For the KdV equation, any weakly nonlocal recursion operator is a polynomial 
of $\Re$ with constant coefficients.
However, this is not true for elliptic models such as the Krichever-Novikov equation
and the Landau-Lifshitz equation. 

In \cite{DS08}, Demskoi and Sokolov showed that the
commutative ring of all weakly nonlocal recursion operators for elliptic
models is isomorphic to the coordinate ring of the elliptic curve. For example,
the Krichever-Novikov equation
$$
u_t=u_{xxx}-\frac{3}{2} \frac{u_{xx}^2}{u_x}+\frac{{\rm P}(u)}{u_x}, 
$$
where ${\rm P}(u)$ is an arbitrary quartic polynomial of dependent variable $u$,
possesses one weakly nonlocal recursion operator $\Re_1$of order $4$
and another one  $\Re_2$ of order $6$. These two operators are related by the
algebraic equation corresponding the elliptic curve associated with the
Krichever-Novikov equation:
\begin{equation}\label{curve1}
 \Re_2^2=\Re_1^3-\phi \Re_1 -\theta,
\end{equation}
where $\phi$ and $\theta$ are proportional to two modular invariants of a
quartic polynomial
${\rm P}(u)$.

In this letter, we first present a new recursion operator valid for both the Viallet equation and
Yamilov's discretisation of the Krichever-Novikov equation. We state that the
new recursion operator and the operator proposed in \cite{mwx1,mwx2} satisfy to
the algebraic equation similar to (\ref{curve1}).

\section{The Viallet equation}
In this section, we give necessary definitions and a short account of relevant
results on our recent study of the symmetries, conservation laws and recursion
operator for the Viallet equation. 

The Viallet equation is a difference equation on $\Z^2$ given by
\begin{eqnarray}
Q &:=& a_1 u_{0,0} u_{1,0} u_{0,1} u_{1,1} \nonumber \\
& & + a_2 (u_{0,0} u_{1,0} u_{0,1} + u_{1,0} u_{0,1} u_{1,1}
 + u_{0,1} u_{1,1} u_{0,0} + u_{1,1} u_{0,0} u_{1,0}) \nonumber\\
&&  + a_3 (u_{0,0} u_{1,0}+u_{0,1} u_{1,1}) +a_4 (u_{1,0} u_{0,1} +
u_{0,0} u_{1,1})  \nonumber\\
&&  + a_5 (u_{0,0} u_{0,1} + u_{1,0} u_{1,1}) + a_6
(u_{0,0}+u_{1,0}+u_{0,1} +u_{1,1})
+ a_7 \,=\,0\,,\label{QV}
\end{eqnarray}
where $a_i$ are arbitrary complex parameters such that the polynomial $Q$ is
irreducible. Here the dependent variable $u$ is a complex-valued function 
of independent variables $n,m\in \Z$. We have two commuting shift maps $\cS$ and $\cT$ defined as
\[\begin{array}{c}
 \cS: u\mapsto u_{1,0}=u(n+1,m),\\ \cT: u\mapsto u_{0,1}=u(n,m+1),\\ \cS^p\cT^q:
 u\mapsto u_{p,q}=u(n+p,m+q). \end{array}
\]
For uniformity of the notation, it is convenient to denote the ``unshifted'' function
$u$ as $u_{0,0}$.

Equation (\ref{QV}) was identified by Viallet with the vanishing of its
algebraic entropy \cite{Viallet}. By a point fractional-linear
transformation,  it can be reduced to Adler's equation, also referred 
as the Q4 equation in the ABS classification \cite{ABS}. In fact, all of the
ABS equations can be obtained from the Viallet equation by a simple
specialisation of parameters.  

With affine linear equation (\ref{QV}), following  \cite{ABS1}, we associate 
a bi-quadratic polynomial  $h$ (the discriminant of $Q$) defined by
\begin{eqnarray}
h(u_{0,0},u_{1,0}) &=&  Q\, \partial_{u_{0,1}} \partial_{u_{1,1}} Q\, -\,
\partial_{u_{0,1}} Q \, \partial_{u_{1,1}} Q \,  \label{h-polynomials-def1}
\end{eqnarray}
and a quartic polynomial $f$ of $u_{0,0}$ given by
\[
 f(u_{0,0})= (\partial_{u_{1,0}} h)^2
-2 h \partial_{u_{1,0}}^2 h .
\]

Affine-linear equation (\ref{QV}) is covariant with respect to the M\"obius
transformations 
$$u_{n,m}\mapsto\frac{\alpha u_{n,m}+\beta}{\gamma
u_{n,m}+\delta},\qquad \alpha\delta-\beta\gamma\ne 0.$$
Classical modular invariants $g_2,g_3\in\C$ are defined as (see \cite{ABS1},
\cite{WhittWat}):
\begin{eqnarray}
&&g_2=\frac{1}{48}\left(2 f f^{IV}-2 f'f'''+(f'')^2\right)\, ,\label{gg2}\\
&&g_3=\frac{1}{3456}\left( 12 f f''f^{IV}-9 (f')^2 f^{IV}-6 f (f''')^2+6 f' f''
f''' -2 (f'')^3\right)\, .\label{gg3}
\end{eqnarray}
Thus with a difference equation defined by an affine-linear polynomial $Q$ one
can associate a plane algebraic curve 
\begin{equation}\label{curve0}
 y^2=4x^3-g_2 x-g_3
\end{equation}
in the Weierstrass form. If the modular discriminant of the curve is non-zero
$g_2^3-27 g_3^2\ne 0$ then the curve (\ref{curve0}) is elliptic
\cite{WhittWat}.

Following  \cite{ABS1} we introduce the
relative invariants $I_2,I_3\in \C$:
\begin{eqnarray}
&&I_2=\frac{1}{6} \left(h \partial_u^2\partial_{u_{1,0}}^2 h-(\partial_u h)
(\partial_u \partial_{u_{1,0}}^2 h)-(\partial_{u_{1,0}} h) (\partial_u^2
\partial_{u_{1,0}} h)
+(\partial_u^2 h) (\partial_{u_{1,0}}^2 h)\right)\\
&&\qquad+\frac{1}{12}(\partial_u \partial_{u_{1,0}} h)^2\, ,\label{II2}\\
&&I_3=\frac{1}{4} \det \left(\begin{array}{ccc}\ h\ &\ \partial_u h\ &\
\partial_u^2 h\ \\
\partial_{u_{1,0}} h & \partial_{u_{1,0}} \partial_u h & \partial_{u_{1,0}}
\partial_u^2 h\\
\partial_{u_{1,0}}^2 h & \partial_{u_{1,0}}^2 \partial_u h &
\partial_{u_{1,0}}^2 \partial_u^2 h
\end{array}\right)\, ,\label{II3}
\end{eqnarray}

We note that the relative and modular invariants are not independent. 
\begin{Pro}
 There is a syzygy among $I_2, I_3, g_2$ and $g_3$, namely
\begin{equation}\label{syzygy}
 I_3^2=4 I_2^3-g_2 I_2-g_3\ .
\end{equation}
\end{Pro}
\noindent
{\bf Proof:} The statement of this proposition can be verified  by a direct
computation. \hfill $\blacksquare$

The Viallet equation (\ref{QV}) is invariant under the involution  
\begin{equation}\label{inv}
u_{1,0}\rightleftarrows u_{0,1},\quad a_3\rightleftarrows a_5.
\end{equation}
This property enables us to study symmetries, recursion operators and
conservation laws only for the one direction of the lattice and to recover the
complementary set for the other direction. 
We denote $\cF_\bs$ the  field of rational functions of variables
$\{u_{n,0}\,|\, n\in\Z\}$. It is a difference field with automorphism $\cS$.

\begin{Def}\label{DefSymmetry} Assume $K\in\cF_\bs$ depends on a finite set
of variables $\{u_{n,0}\,|\, n_1\leq n\leq n_2\}$ with $\partial_{u_{n_1,0}} K
\neq 0$ and $\partial_{u_{n_2,0}} K \neq 0$.
We say $K$ is  a {\em symmetry} of equation (\ref{QV}) of order $(n_1,n_2)$ if
$$D_Q(K)=0\quad \mbox{for all solutions of}\quad Q=0.$$
Here $D_Q$ is the Fr\'echet derivative of $Q$ defined as
\begin{equation}\label{frechet}
D_Q=\frac{\partial Q}{\partial u_{1,1}} \ \cS \cT +\frac{\partial Q}{\partial
u_{0,1}}  \cT 
+\frac{\partial Q}{\partial u_{1,0}} \ \cS+\frac{\partial Q}{\partial u_{0,0}}\ .
\end{equation}
\end{Def}

Equation (\ref{QV}) possesses a
generalised symmetry of order $(-1,1)$
\cite{TTX, X, mwx1}:
\begin{eqnarray}
&& K^{(1)} = \frac{h}{u_{1,0}-u_{-1,0}} - \frac{1}{2} \partial_{u_{1,0}} h\,
. \label{K1}
\end{eqnarray}
The corresponding  generalised symmetry for equation (\ref{QV}) in
$\cT$ direction is
\begin{eqnarray*}
&& \frac{{\hat h}}{u_{0,1}-u_{0,-1}} - \frac{1}{2}
\partial_{u_{0,1}} {\hat h}\, 
\end{eqnarray*}
with the polynomial ${\hat h}=Q\, \partial_{u_{1,0}} \partial_{u_{1,1}} Q\, -\,
\partial_{u_{1,0}} Q \, \partial_{u_{1,1}} Q$ . It is related with (\ref{K1})
by the involution (\ref{inv}).

For convenience from now on, we will use one-index
notation to denote shifts of elements $\cF_\bs$ by $\cS$, e.g. $K^{(1)}_{j}=\cS^j ( K^{(1)})$ 
and $h_n=\cS^n(h(u_{0,0},u_{1,0}))$. We shall omit the index zero for
unshifted functions. Often we shall use notation  
\begin{equation}
 w=\frac{1}{u_{1,0}-u_{-1,0}} \quad \mbox{and} \quad  w_k=\cS^k w. \label{w}
\end{equation}

In this notations a generalised symmetries of the Viallet equation of 
order $(-2,2)$ can be written as \cite{TTX, X, mr89k:58132, mwx1}:
\begin{equation}
K^{(2)} = h\,h_{-1}w^2(w_1+w_{-1}) \,.\label{K2}
\end{equation}

In our recent work \cite{mwx1,mwx2}, we proved that infinitely many generalised 
symmetries $K^{(n)}, \ \mbox{ord}K^{(n)}=(-n,n)$ of the Viallet equation (\ref{QV})
can be generated by a recursion operator $\gR$
\[
 K^{(2n-1)}=\gR^{n-1}(K^{(1)}),\qquad  K^{(2n)}=\gR^{n-1}(K^{(2)}),\qquad
n\in\N .
\]
Here the recursion operator $\gR$ is a second order  pseudo-difference operator
of the form
\begin{eqnarray}
\gR &=& h\, h_{-1}\,w^2 w_{1}^2 \cS^2 + h\, h_{-1}\,w^2 w_{-1}^2\cS^{-2}
+2 K^{(1)} K^{(2)} \left(\frac{1}{h} \cS+\cS^{-1}\frac{1}{h}\right) \nonumber \\
&& - w^2 \left(h_{-1}\, h_{1} w_{1}^2+ h_{-2}\, h\,w_{-1}^2\right)
+\frac{2}{h_{-1}} \left( K^{(1)} K^{(2)}_{-1} + K^{(2)} K^{(1)}_{-1} \right) \nonumber \\
& & +\, 2\,\,K^{(1)}\, (\cS-1)^{-1}\circ
\left(\frac{K^{(2)}_{-1}}{h_{-1}}\,-\frac{K^{(2)}_{1}}{h} \right) 
+ 2\,\,K^{(2)}\, (\cS-1)^{-1}\circ 
\left(\frac{K^{(1)}_{-1}}{h_{-1}}\,-\frac{K^{(1)}_{1}}{h} \right)\, 
\label{weakR}.
\end{eqnarray} 

For instance, $\gR(K^{(1)})$ gives rise of the symmetry of order $(-3,3)$ for
equation
(\ref{QV}) \cite{mwx1} as follows:
\begin{eqnarray}
K^{(3)}=\gR(K^{(1)})= h\, h_{-1} w^2
\left(K^{(1)}_{2} w_1^2+ K^{(1)}_{-2} w_{-1}^2\right) +\left(w_1 + w_{-1}
\right) K^{(1)} K^{(2)} \,.\label{K3}
\end{eqnarray}

Operator $\gR$ is a sum of a difference operator with coefficients in
$\cF_{\bs}$ and a non-local pseudo-difference operator.
The nonlocal part of the operator $\gR$ (\ref{weakR}) is a
finite sum of the form $a\ (\cS-1)^{-1}\circ b$
with $a,b\in \cF_{\bs}$. Such operators belong to the type of weakly nonlocal
pseudo-difference operators \cite{mwx1, mwx2}. 
It is a difference analogue of weakly nonlocal pseudo-differential operators
studied by Maltsev and Novikov \cite{mr2002g:37093}. 

Recursion operator $\gR$ can be presented as the product $\gR={\cal H}\circ \J$
of the Hamiltonian operator ${\cal H}$  and the symplectic operator $\J$,  where
\begin{eqnarray}
{\cal{H}} &=& h_{-1}\, h\, h_{1} w^2 w_1^2\, {\cal{S}}\,-\,
{\cal{S}}^{-1} h_{-1}\, h\, h_{1} w^2 w_1^2+\, 2\,K^{(1)}\,  (\cS-1)^{-1}{\cal{S}} \circ K^{(2)}\,
\label{hamilt-gen}\\
&&\quad +\, 2\, K^{(2)}\, (\cS-1)^{-1}\circ K^{(1)} \,,\nonumber \\
\J&=& \frac{1}{h}\ \cS -\cS^{-1}\ \frac{1}{h} \,. \label{symp-gen}
\end{eqnarray}
The Hamiltonian operator ${\cal H}$ 
maps the variational derivatives of
conserved densities to symmetries while the symplectic operator $\J$
maps
symmetries to the variational derivatives of conserved densities
\cite{mwx1,mwx2}.

\section{A new recursion operator for the Viallet equation}
In this section, we present  new weakly nonlocal recursion and
Hamiltonian operators for the Viallet equation (\ref{QV}). The new recursion
operator and the recursion operator (\ref{weakR}) satisfy to the agebraic
equation corresponding to the elliptic curve associated with the Viallet
equation. It is a product of the new 
Hamiltonian operator and the symplectic operator $\J$ given by (\ref{symp-gen}).
\begin{The}
A weakly nonlocal pseudo-difference operator
\begin{eqnarray}
{\hat \gR}&=&\frac{A}{h_2} \cS^3+\frac{B}{h_1}\cS^2 
+\left(\frac{{K^{(2)}}^2}{h}+\frac{2 K^{(1)} K^{(3)}}{h} -\frac{A}{h_1}\right)\cS
+\frac{2}{h_{-1}}\left(K^{(2)} K^{(2)}_{-1} + K^{(1)} K^{(3)}_{-1}+K^{(3)} K^{(1)}_{-1}\right)\nonumber\\
&& - \frac{B}{h}-\frac{B_{-1}}{h_{-1}}+\left(\frac{{K^{(2)}}^2}{h_{-1}}+\frac{2 K^{(1)} K^{(3)}}{h_{-1}} -\frac{A_{-2}}{h_{-2}}\right)\cS^{-1}
+\frac{B_{-1}}{h_{-2}}\cS^{-2}+\frac{A_{-2}}{h_{-3}} \cS^{-3}\nonumber\\
&&+\, 2\,\,K^{(1)}\, (\cS-1)^{-1}\circ
\left(\frac{K^{(3)}_{-1}}{h_{-1}}\,-\frac{K^{(3)}_{1}}{h} \right) 
+ 2\,\,K^{(3)}\, (\cS-1)^{-1}\circ 
\left(\frac{K^{(1)}_{-1}}{h_{-1}}\,-\frac{K^{(1)}_{1}}{h} \right)\nonumber\\
&&+2\,\,K^{(2)}\, (\cS-1)^{-1}\circ
\left(\frac{K^{(2)}_{-1}}{h_{-1}}\,-\frac{K^{(2)}_{1}}{h} \right)\, ,
\label{weak2R}
\end{eqnarray}
where $w_i$ is defined by (\ref{w}); $h$ is the discriminant of $Q$ given by
(\ref{h-polynomials-def1}),  $K^{(j)}, j=1,2,3$ are symetries of equation
(\ref{QV}) given by (\ref{K1}), (\ref{K2}) and (\ref{K3}), and 
\begin{eqnarray}
A&=& h_{-1}\, h\, h_{1} h_2 w^2 w_1^2 w_2^2\ ;\label{coefa}\\
B&=& 2 h_{-1} w^2 \left(  h h_1 w_1^2 (K^{(1)} w_{-1} -\frac{1}{2} w
\partial_{u_{0,0}} h +\frac{1}{4} \partial_{u_{0,0}} \partial_{u_{1,0}} h ) +
K^{(1)}_1 K^{(2)}_1\right)\ ,\label{coefb}
\end{eqnarray}
is a recursion operator for equation (\ref{QV}).
\end{The}
In the proof that $\gR$ (\ref{weakR}) is a recursion operator of equation
(\ref{QV}), cf. Theorem 1 in \cite{mwx2},
the key observation was that there exists a constant $\mu$
\begin{eqnarray*}
&&\mu=(a_4-a_3)(a_3 a_1 a_7-a_3 a_5^2-2 a_7 a_2^2+4 a_5 a_2 a_6-2 a_1 a_6^2+a_4
a_1 a_7-a_4 a_5^2)\ , 
\end{eqnarray*}
such that the operator  $\gR-\mu$ can be represented in the factorised form
\begin{eqnarray*}
\gR-\mu &=&\cM \cdot \left( \frac{\partial Q}{\partial u_{1,0}} \cS+
\frac{\partial Q}{\partial u_{0,0}}\right),
\end{eqnarray*}
where $\cM$ is a weakly non-local pseudo-difference operator, explicitly given
in Lemma 2 in \cite{mwx2}. In the case of the operator $\hat{\gR}$
(\ref{weak2R})
one can check that for
\begin{eqnarray*}
 \hat{\mu}=I_3+\frac{1}{2}\mu (a_1 a_7+a_3^2+a_4^2+a_5^2-4 a_2 a_6)
\end{eqnarray*}
the operator  $\hat{\gR}-\hat{\mu}$ also can be represented in the factorised
form 
\begin{eqnarray*}
\hat{\gR}-\hat{\mu} &=&\hat{\cM} \cdot \left( \frac{\partial Q}{\partial
u_{1,0}} \cS+
\frac{\partial Q}{\partial u_{0,0}}\right),
\end{eqnarray*}
with a certain weakly non-local pseudo-difference operator $\hat{\cM}$ (the
explicit form of $\hat{\cM}$ is rather long and we omit it in this short paper).
The rest of the proof of this theorem is not different from the proof of Theorem
1 in
\cite{mwx2} which states that $\gR$ is a
recursion
operator for the Viallet equation.

\begin{Pro} The recursion operator
${\hat \gR}$ can be written as the product ${\hat \gR}={\hat \cH}\circ \J$ of
the Hamiltonian operator ${\hat \cH}$
and the symplectic operator $\J$ given by (\ref{symp-gen}), where
\begin{eqnarray}
{\hat {\cal{H}}} &=& A\, {\cal{S}}^2\,-\,
{\cal{S}}^{-2} A +B \cS-\cS^{-1} B +\, K^{(2)}\,  (\cS-1)^{-1}({\cal{S}}+1) \circ K^{(2)} \nonumber \\
& &  +\, 2\,K^{(1)}\,  (\cS-1)^{-1}{\cal{S}} \circ K^{(3)}\, +\, 2\, K^{(3)}\,
(\cS-1)^{-1}\circ K^{(1)} \,, \label{hamilt2-gen}
\end{eqnarray}
and $A$ and $B$ are defined in (\ref{coefa}) and (\ref{coefb}).
\end{Pro}

Computing the product ${\hat \cH}\circ \J$ one can directly verify that the
result coincides with ${\hat \gR}$ (\ref{weak2R}). Operator ${\hat {\cal{H}}}$
is obviously skew-symmetric. The proof of the Jacobi identity for the Poisson
bracket corresponding to ${\hat {\cal{H}}}$ is similar to one for the operator
${{\cal{H}}}$ (Proposition 2 in \cite{mwx2}).

The central result of this paper is formulated in the following Theorem.

\begin{The}\label{th1} {\rm (i)} The recursion operators $\gR$ (\ref{weakR})
and
$\hat{\gR}$ (\ref{weak2R}) satisfy the algebraic equation
\begin{eqnarray}
 (2 {\hat \gR} -I_3)^2=4 (\gR+I_2)^3- g_2 (\gR+I_2)-g_3, \label{relation}
\end{eqnarray}
where $I_2, I_3, g_2$ and $g_3$ are the relative and modular invariants
(\ref{II2}), (\ref{II3}), (\ref{gg2}) and (\ref{gg3}).\\ {\rm (ii)}
Operators $\gR$ and $\hat{\gR}$ commute.
\end{The}

Every pseudo-difference operator can be uniquely represented by its formal
Laurent series and there is a unique formal series corresponding to a formal
inverse of the operator \cite{mwx1}. For example operator $\gR$ can be
represented by 
\[
 \gR_L= h\, h_{-1}\,w^2 w_{1}^2 \cS^2 
+\frac{2}{h} K^{(1)} K^{(2)}  \cS+\cdots\, ,\]
and its formal inverse by
\[ \gR_L^{-1}=
\frac{1}{h_{-2}\, h_{-3}\,w_{-2}^2 w_{-1}^2} \cS^{-2}-\frac{2  K^{(1)}_{-2}
K^{(2)}_{-2}}{h_{-2}^2\, h_{-3}^2\,h_{-4}\,w^2_{-1} w_{-2}^4 w_{-3}^2} 
 \cS^{-3}+\cdots\, .
\]
It follows from Theorem 1, that equation (\ref{QV}) has the first order
formal recursion operator 
\[\tilde{\gR}=\hat{\gR}_L\circ \gR_L^{-1}=h_{-1}w^2\cS+\cdots \, . 
\]
Logarithmic residue $\rho_0=\mbox{res\,}\log\tilde{\gR}=\log(h_{-1}w^2)$ and 
$\rho_k=\mbox{res}\,\tilde{\gR}^k,\ k=1,2,3,\ldots $ are densities of canonical
conservation laws for the difference equation (\ref{QV}) (see Theorem 4 in
\cite{mwx1}).

In \cite{mwx1} we claimed that we can compute a square root of the recursion
operator $\gR$ (\ref{weakR}), although for a general second order formal series
a square root may not exist.
\begin{Pro}
 There exist  first order formal series $R=h_{-1}w^2\cS+r_0+r_1 \cS^{-1}+\cdots
$ and
$\hat{R}=h_{-1}w^2\cS+\hat{r}_0+\hat{r}_1 \cS^{-1}+\cdots $ with
coefficients  $r_k,\hat{r}_k\in\cF_\bs$ such that $\gR_L=R^2$ and
$\hat{\gR}_L=\hat{R}^3$. 
\end{Pro}
{\bf Proof:}
The existence of the square root of $\gR$
(\ref{weakR}) and a cubic root of $\hat{\gR}$ (\ref{weak2R}) follow from Theorem
\ref{th1} and the following obvious Lemma:
\begin{Lem}\label{lem1}
 Let $\cA$ and $\cB$ be two commuting formal series and ${\rm ord}\,\cA^k>{\rm
ord}\,\cB$ for an integrer $k>0$. Then the $k$-th root
$\cC=(\cA^k+\cB)^{\frac{1}{k}}$ is given by the formal series
$\cC=\cA+\sum_{j>0}C(\frac{1}{k},j)\cB^j \cA^{1-jk}$, where
$C(\frac{1}{k},j)=\frac{1}{j!}\prod_{i=1}^j (\frac{1}{k}-i+1)$ are
generalised binomial coefficients.
\end{Lem}
Indeed, after re-arranging equation (\ref{relation}) as
\[
 \gR=\frac{1}{4}(2 \hat{\gR}-I_3)^2 ({\gR}+I_2)^{-2}-I_2
+\frac{g_2}{4}({\gR}+I_2)^{-1}+\frac{g_3}{4}({\gR}+I_2)^{-2}
\]
we can apply Lemma \ref{lem1} with 
\[k=2,\quad \cA=\frac{1}{2}(2 \hat{\gR}_L-I_3)
({\gR}_L+I_2)^{-1},\quad \cB= -I_2
+\frac{g_2}{4}({\gR}_L+I_2)^{-1}+\frac{g_3}{4}({\gR}_L+I_2)^{-2}.
\]
We have ${\rm ord}\,
\cA=1,\ {\rm ord}\, \cB\le 0$ and $[\gR,\hat{\gR}]=0$ implies  $[\cA,\cB]=0$.
Then $R=\cC$ is a first order formal recursion operator with
coefficients in $\cF_\bs$, such that  $\gR_L=R^2$. In a similar way one can find
 $\hat{R}, \ $ such that  $\hat{\gR}_L=\hat{R}^3$.\hfill $\blacksquare$

\section{Yamilov's discretisation of the Krichever-Novikov equation}
 As pointed in \cite{mwx2}, the operator $\gR$ (\ref{weakR}) is a recursion
operator of differential-difference equation $u_{t_1}=K^{(1)}$, which can be
identified as Yamilov's discretization of the Krichever-Novikov equation (YdKN)
\cite{Yami1}, cf. equation (V4) when $\nu=0$ in \cite{Yami}. Indeed
\begin{equation}
u_{t_1}=K^{(1)}=\frac{R(u_{1,0},u,u_{-1,0})}{u_{1,0}-u_{-1,0}}, \label{yamV4}
\end{equation}
where
$$
R(u_{1,0},u,u_{-1,0})=( b_1 u^2+ 2 b_2 u+b_3) u_{1,0} u_{-1,0} + ( b_2 u^2+ b_4 u +b_5) (u_{1,0}+u_{-1,0}) + b_3 u^2 +2 b_5 u +b_6\ .
$$
and 
\begin{eqnarray*}\label{para}
\begin{array}{lll} b_1=a_1 a_3-a_2^2; & b_2=
{\frac{1}{2}} 
(a_2 a_3 +a_1 a_6-a_2
a_5-a_2 a_4); & b_3= a_2 a_6-a_4 a_5;\\ &&\\
b_4=\frac{1}{2}(a_3^2+a_1 a_7-a_4^2-a_5^2); & b_5=\frac{1}{2}(a_3 a_6+a_2
a_7-a_4 a_6-a_5 a_6); & b_6=a_3 a_7-a_6^2.
 \end{array}
\end{eqnarray*}
It is straightforward to check that
\begin{eqnarray}\label{hR}
 h(u,u_{1,0}) =R(u_{1,0},u, u_{1,0}).
\end{eqnarray}
Using (\ref{yamV4}),(\ref{hR}) we can express $\gR$ (\ref{weakR}), $\cH$
(\ref{hamilt-gen}) and $\J$ (\ref{symp-gen}) in terms of the polynomial
$R(u_{1,0},u,u_{-1,0})$. In \cite{mwx2} it has been shown that  
$\cH$, $\J$ and $\gR$ are Hamiltonian, symplectic and recursion
operators for the YdKN equation respectively.
  By direct computation,
we can show that 
$$L_{K^{(1)}}{\hat \cH}=0,$$
where $L_{K^{(1)}}$ is the Lie derivative in the direction of the vector field
$K^{(1)}$ (for a detail definition of the Lie derivative see \cite{mwx2}). Thus
${\hat \cH}$ is a Hamiltonian operator for
equation (\ref{yamV4}) and $\hat \gR$ is a recursion operator of YdKN
(\ref{yamV4}). The mentioned above canonical conserved densities $\rho_k, \
k=0,1,2,\ldots$ are densities of local conservation laws of the YdKN equation.



\end{document}